\documentclass{ws-procs975x65}

\begin{document}


\newcommand{\CC}{\mathbb C}
\newcommand{\C}{\mathcal C}
\newcommand{\RR}{\mathbb R}
\newcommand{\ZZ}{\mathbb Z}
\newcommand{\NN}{\mathbb N}
\def\d{\partial}
\def\p{\varrho}
\def\w{\omega}
\def\Vac{\Omega}

\def\eps{\epsilon}

\def\v#1{\vec{#1}}

\def\a{\alpha}
\def\be{\beta}
\def\A{\Lambda}
\def\w{\omega}
\def\Vac{\Omega}
\def\ro{\varrho}
\def\e{\epsilon}
\def\ve{\varepsilon}
\def\s{\sigma}
\def\g{\gamma}
\def\G{\Gamma}
\def\de{\delta}
\def\De{\Delta}
\def\m{\mu}
\def\n{\nu}
\def\vp{\varphi}
\def\la{\lambda}
\def\La{\Lambda}

\def\h{{\mathsf h}}
\def\H{\mathcal H}
\def\Rup{\overrightarrow{\mathsf R}}
\def\Rin{\overleftarrow{\mathsf R}}
\def\Bin{\overleftarrow{\mathsf B}}
\def\Rpl{{\mathsf R}_{lp}}
\def\Rwl{{\mathsf R}_{l\w}}
\def\R{r_s}
\def\Hb{\mathcal H}
\def\dens#1{\langle\varrho(#1)\rangle}
\def\r{{r^*}}
\def\V{\mathcal V}

\newsymbol\rest 1316

\newcommand{\ca}[1]{{\cal #1}}

\def\d{\partial}
\def\D{\nabla}
\def\dd#1#2{\frac{\de #1}{\de #2}}
\def\DD#1#2{\frac{\d #1}{\d #2}}
\def\OpD{\mathcal{D}}

\def\con#1{#1^\dagger}
\def\b#1{\overline{#1}}
\def\T#1{\tilde{#1}}
\def\*{\cdot }
\def\nin{\in \hspace{-3.5mm} /}
\def\dir#1{#1 \hspace{-2.6mm} \slash}
\def\del#1{\widetilde{#1}}
\def\ra{\rightarrow}
\def\rest{\upharpoonright}

\renewcommand{\Im}{\text{Im}}
\renewcommand{\Re}{\text{Re}}

\def\K{\mathcal{K}}
\def\Fk{\Fou_\K}
\def\C{\mathcal{C}}
\def\Dom{\mathcal{D}}
\def\Dz{\mathcal{D}}
\def\M{\mathcal{M}}
\def\Fou{\mathcal{F}}
\def\A{\mathcal{A}}
\def\W{\mathcal{W}}
\def\Lag{\mathcal{L}}
\def\Alg{\hbox{$O\hspace{-2.2pt}\iota
\hspace{-2.2pt}^{\prime}\hspace{-3.1pt}\acute{}\hspace{3pt}$}}

\def\O{\mathcal{O}}

\def\ess_sp{\s_{ess}}
\def\F1{\Fou^{(1)}}
\def\PDO{$\Psi$DO}

\def\Z{\mathbb{Z}}
\def\N{\mathbb{N}}
\def\U{\mathbb{U}}
\def\S{\mathbb{S}}
\def\id{\mathbf{1}} 

\def\ddd#1{d^3#1\ }
\def\dm#1{\frac{d^3#1}{2 #1_0}\ }

\def\C{\mathcal C}

\def\dom{\C_0^\infty}
\def\smooth{\C^\infty}
\def\lsq{L^2(\R^3,d^3x)}


\def\sp{\vspace{1em}}
\def\M{\mathfrak M}

\def\t#1{\mbox{#1}}


\title{ON QUANTUM EFFECTS IN THE VICINITY OF WOULD-BE HORIZONS\footnote{This research has been supported by the Kosicuszko Foundation}}

\author{PIOTR MARECKI}

\address{ITP, Universit\"at Leipzig\\
D-04009 Leipzig, Germany 
\email{pmarecki@gmail.com}}


\begin{abstract}
 We present a method based on the so-called Quantum Energy Inequalities, which allows to compare, and bound, the expectation values of energy-densities of ground states of quantum fields in spacetimes possessing isometric regions. The method supports the conclusion, that the Boulware energy density is universal both: at modest (and far) distances from compact spherical objects, and close to the would-be horizons of the gravastar/QBHO spacetimes. It also provides a natural consistency check for concrete (approximate, numerical) calculations of the expectation values of the energy-momentum tensors. 
\end{abstract}

\bodymatter

\section{Introduction}\label{intro}

	One of the most important aspects of Quantum Field Theory in curved spacetimes is the existence of non-vanishing expectation values of the energy-momentum-tensor operator in virtually all states of the quantum fields. Yet in concrete cases, for specific spacetimes and states believed to be of physical interest, the calculation of these expectation values is notoriously difficult. Not only the classical solutions of the wave equation, $F_I(\vec x,t)$, need to be known\footnote{Which already exceeds our capabilities, for instance, in the Schwarzschild spacetime}, but also sums of (absolute) squares of these solutions over their index $I$ need to be known at every point. Only in few precious cases has this been possible, often in approximate sense only. Here we report on a method, which allows to use the result known for one spacetime to approximate the energy densities on another spacetime, which is isometric to the initial one in some region only. As an application, the known properties of the Boulware state in the Schwarzschild geometry provide a bound on the energy densities of ground states in spacetimes of compact, spherically symmetric objects. 

Let us take the spherically symmetric spacetimes, with the (general) metric
\begin{equation}
  ds^2=f(r) dt^2-\frac{dr^2}{h(r)} -r^2 (d\theta^2 +\sin^2\theta d\phi),
\end{equation}
with the suitable functions $f$ and $h$.  The main object of interest is the (temporally smeared) energy density operator, 
\begin{equation}\label{ro}
  \rho(w,\vec x)= \int dt\ T_{ab}(\vec x,t)\ u^a u^b\, w^2(t).
\end{equation}
where the operator $T_{ab}$ has been constructed by point-splitting w.r.t. a Hadamard parametrix\footnote{We mean the standard construction employed in most concrete calculations, and note, that such an operator is locally covariant \cite{MA,FP}.}. The function, $w(t)$ is smooth, compactly supported, and normalized as a probability density, $\int dt\ w^2(t) =1$, while $u^a$ denotes the tangent vector of the observer measuring the energy density. 

We shall employ the method of Quantum Energy Inequalities \cite{TR,PF,FP}, which asserts, that \emph{all} Hadamard states $\psi$ must necessarily fulfill
\begin{equation}\label{QEI}
   \langle\p(w,\vec x)\rangle_\psi - \langle \p(w,\vec x)\rangle_G
\geqslant -Q_G[w,\vec x],
\end{equation}
 where $G$ indicates the ground state of the quantum field, and
\begin{equation}\label{arb_reference}
Q_G=\frac{1}{\pi}\int_0^\infty
d\w\int_{-\infty}^{\infty} dt\,ds\, e^{-i\w(t-s)} w(t)w(s)
\left[\frac{\d_t\d_s}{f(r)} -(\D^x)_i(\D^y)^i\right] \w^G_2(t,\vec
x,s,\vec y)|_{\vec y=\vec x}.
\end{equation}
Here, $\w^G_2(t,\vec x,s,\vec y)$ stands for the two-point
function of the ground state\footnote{The QEI are proved to be true also with respect to any other Hadamard state \cite{FP}, that is, we can replace the ground state $G$ by any other Hadamard state (although such a replacement changes also the right-hand side of the inequality).}. 
The functionals $Q_{G}[w,\vec x]$ can conveniently be expressed by the respective mode-sums:
\begin{multline}\label{QEIss} Q_G[w,\vec x]=
\frac{1}{16\pi^3}\int_0^\infty
  d\w\int_0^\infty \frac{dp}{p}
  \left[\frac{p^2}{f}-\frac{H}{r^2}\d_r(H r^2\d_r)
  \right] \left[\sum_{l=0}^\infty (2l+1) |\Rpl(r)|^2\right]\*g(\w+p),
\end{multline}
where $F_I(\vec x)=\frac{1}{\sqrt{2\pi }} \Rwl(r)\ Y_{lm}(\theta,\phi)$ are the solutions of the wave equation corresponding to the frequency $\w$, and $g(\w)=\left| \int_{-\infty}^\infty dt\, w(t) e^{i\w t} \right|^2$. Here also $H(r)=\sqrt{fh}$, is derived from the metric functions.

The functional $Q_G$, which is finite and always positive, is independent of the quantum state $\psi$, which is remarkable, as the variety of quantum states is enormous. Even more surprising are the (typical) properties of $Q_G$: its value falls off rapidly as the time of measurement gets longer, i.e. as the support of $w(t)$ gets larger. Thus, substantial sub-ground-state  energy densities can be produced by quantum fields only for short durations of time, and it is impossible to keep these densities sub-ground for longer time periods.

\section{Energy-densities in locally isometric spacetimes}
 Suppose that two static spacetimes are isometric in a region outside of a sphere at $r=R$. 
If we focus on operators supported in the isometric region, than  we can establish two QEI's \cite{FP,MA}: firstly with $G1$ as the reference state (with $G2$ playing the role of $\psi$), and secondly with the exchanged role of the states. This provides bounds from above and below on the difference of the energy densities:
\begin{equation}\label{QEI1}
   Q_{G1}[w,\vec x]\geqslant \langle\p(\vec x)\rangle_{G1} - \langle \p(\vec x)\rangle_{G2}
\geqslant -Q_{G2}[w,\vec x].
\end{equation}
If we were able to streach the support of $w(t)$ arbitrarily, then the above inequalities would force the difference $\langle\p(\vec x)\rangle_{G2} - \langle \p(\vec x)\rangle_{G1}$ to vanish!
However, there is a QFT subtelty: a priori the states denoted by $\psi$ in \eqref{QEI} should be expressible as density matrices in the Fock space constructed upon $G$. In no way is this possible for $\psi=G2$ and $G=G1$, as these are states providing expectation values of observables from completely incompatible algebras. Nonetheless, the inequalities hold (with expectation values calculated in the usual way) if the smallest double cone containing the support of the operators (in this case: the segment of the curve on which we measure the energy-density) does remain in the isometric region\footnote{In other words: all causal curves, with their ends attached to events of the beginning and the end of the energy measurement, should not leave the isometric region.}.

Thus, the distance between $\vec x$ and the boundary of the isometric region limits the measurement duration\footnote{The maximal duration, $T$, is $\vec x$ dependent, and grows together with the distance from the boundary, so that the bounds \eqref{QEI1} get tighter.}, $T$, and allows a non-vanishing $\langle\p(\vec x)\rangle_{G2} - \langle \p(\vec x)\rangle_{G1}$.

\section{Remarks and outlook}
The bounds \eqref{QEI1} can be directly employed as a new consistency check to be fulfilled in concrete calculations of energy-momentum tensors of quantum fields\cite{A,JMO}.
 They are especially tight if the allowed times of measurement are long, which happens either: far away  (in terms of the geodesic distance) from the boundary of the isometric region, or if the boundary is close to an horizon. In these cases, the functionals $Q_G[w,\vec x]$ depend only on the low frequency asymptotics of the mode-sums. In turn this hints at universality of certain results of QFT in curved spacetimes, notably, our bounds indicate\cite{MA} that the energy density of the Boulware state is universal both: far away from the horizon (always) and close to the horizon, for spacetimes isometric with Schwarzschild up to small distances from the horizon, such as the gravastar or QBHO spacetimes\cite{MM,CL}.

\section*{Acknowledgments}
The author would like to thank Professor 
Pawel O. Mazur for the opportunity to present this research at the BHT5 section of the 11th Marcel Grossman Meeting. The financial support of the 
Kosciuszko Foundation is also gratefully acknowledged.

\vfill


\begin{thebibliography}{00}


\bibitem{MA} P. Marecki, {\it Phys. Rev.} {\bf D73}, 124009 (2006); gr-qc/0507089.

\bibitem{FP} C.J. Fewster and M.J. Pfenning {\it J. Math. Phys. } {\bf 47}, 082303 (2006).

\bibitem{TR} T. A. Roman {\it Proc. 10th Marcel Grossmann Meeting}; gr-qc/0409090.

\bibitem{PF} M.J. Pfenning and L. H. Ford  {\it Phys. Rev. } {\bf D57}, 3489 (1998).


\bibitem{A} P. R. Anderson, W.A. Hiscock, and D.A. Samuel {\it Phys. Rev.} {\bf  D51}, 4337 (1995). 

\bibitem{JMO} J. G. Jensen, B.P. McLaughlin and A.C. Ottewill {\it Phys. Rev.} {\bf D45}, 3002 (1992). 

\bibitem{CL} G. Chapline, et al. {\it Int. J. Mod. Phys.} {\bf A18}, 3587 (2003).

\bibitem{MM} P. O. Mazur and E. Mottola {\it Proc. Nat. Acad. Sci.} {\bf 111}, 9545 (2004). 



\end{thebibliography}
\end{document}